\documentclass{aastex}
\usepackage{spr-astr-addons}
\usepackage{latexsym}
\usepackage{epsfig} 
\usepackage{natbib}
\usepackage{graphicx}
\usepackage{amssymb,amsbsy}
\usepackage{amsmath}
 
\RequirePackage{color}
\newcommand{\msun}{$M_\odot$}

\def\fr{\frac}

\def\half{\frac{1}{2}}

\newcommand{\barr}{\begin{array}}
\newcommand{\earr}{\end{array}}
\newcommand{\berr}{\begin{eqnarray}}
\newcommand{\err}{\end{eqnarray}}
\newcommand{\berrno}{\begin{eqnarray*}}
\newcommand{\errno}{\end{eqnarray*}}
\newcommand{\be}{\begin{equation}}

\newcommand{\gv}{gravitational\,\,}

\newcommand{\dder}[2]{\frac{d #1}{d #2 }}

\newcommand{\pol}[1]{\stackrel{\rm LCP}{\mathrm{RCP}}}

\newcommand{\me}{\mathrm{e}}

\renewcommand{\a}{\alpha}

\newcommand{\g}{\gamma}

\renewcommand{\d}{\delta}

\newcommand{\e}{\epsilon}
\renewcommand{\l}{\lambda}

\newcommand{\s}{\sigma}

\newcommand{\T}{TAUVEX\,\,}
\def\ns{\!\!}

\shorttitle{TAUVEX: UV Transients}
\shortauthors{Safonova, Sivaram, and Murthy}
\begin{document}

\title{Prospects for  Observations of Transient UV Events 
with the TAUVEX UV Observatory}
 
\author{Margarita Safonova\altaffilmark{1}}
\and
\author{C.~Sivaram\altaffilmark{2}}
\and
\author{Jayant Murthy\altaffilmark{3}}
\affil{Indian Institute of Astrophysics, Koramangala 4th block,
Bangalore 560012, India}
 
\altaffiltext{1}{rita@iiap.res.in}
\altaffiltext{2}{murthy@iiap.res.in}
\altaffiltext{3}{sivaram@iiap.res.in}

\begin{abstract}
Transient events have posed special problems in astronomy because of the 
intrinsic difficulty of their detection, and a new class of observatories 
such as the Pan-STARRS and LSST are coming up specifically to observe these energetic events.
In this paper we discuss the UV transient events from two specific sources, such as
possible collisions in extrasolar planetary systems and M dwarf flares, 
to find the probability of their detection by space UV observatories, in particular, 
by the Tel Aviv University Explorer (TAUVEX). TAUVEX is an UV 
imaging experiment that will image large parts of the sky in the wavelength 
region between 1200 and 3500 \AA. \T is a collaborative effort between the 
Indian Institute of Astrophysics (IIA) and Tel Aviv University, and is scheduled for an 
early-2009 launch with at least three years of operations. The scientific instrument 
has been fabricated at El-Op in Israel, with the satellite interfaces, launch and flight 
operations provided by the Indian Space Research Organization (ISRO). The ground-based
software development is the responsibility of the IIA 
while other aspects of the mission are the joint responsibility of IIA and
Tel Aviv University. \T Science Team (TST) have 
created a coherent observing program to address several key science objectives,
one of them is a program to study short-scale UV transient events. 
We have estimated that in one year of TAUVEX observations we can expect about 
$90-350$ short-scale transient events. Because we obtain real-time telemetry 
with TAUVEX, we will be able to catch transients early in their evolution and 
to alert other observatories. We also present a description of \T mission, 
including instrument design and its estimated performance.

\end{abstract}

\keywords{TAUVEX: --- space missions: planetary collisions; UV flares}

\section{Introduction}

Astrophysical transient events include bursts, flashes, flares, 
gamma-ray bursts (GRBs), supernovae (SN), etc. The discovery and 
study of highly transient sources, especially those that rise to 
high brightness and then fade, has always been a major part 
of modern astrophysics. Historically, they were studied either 
as very high-energy phenomena, or in infrared (IR). 
For example, even before the discovery of extra-solar planets (ESPs) around main 
sequence (MS) stars, Stern (1994) discussed the possibility of detection 
of planets through IR signals from collisions with their parent stars. 
Recently, it was proposed that planet-planet 
collisions will give rise to flares, where at the flash peak time, the extreme-UV (EUV) 
flash can greatly outshine the host star \citep{zhang}. On the observational side, 
the last two years (2006 and 2007) have been witness to the 
importance of the UV space survey 
missions in detection of UV flares \citep{galex-flares,galex-uvceti-flash}. 
The conclusion derived from the Deep Lens Survey (DLS) that late-type dwarf flares 
constitute a dense ``foreground fog" in our Galaxy, hindering any extragalactic transient
search, brings back a question on a derivation of the flare frequency rate, 
with eventual identification of all nearby M dwarfs \citep{DLS}.

With the launch of \T on board the Indian geostationary satellite 
GSAT-4, the astronomical community 
will have access to a flexible instrument for observations in the mid-UV 
range of $135-400$ nm. \T is expected to observe to a greater depth than the
NASA Galaxy Explorer satellite ({\it GALEX}) \cite{Martin2005} in $20\%$ of the
sky and in three simultaneous UV bands compared to two of {\it GALEX}. 
The three parallel telescopes of \T will enable to image simultaneously at three
different frequency bands, thus measuring UV colours of all objects in the
image frame. In parts of the orbit with little or 
no straylight the detection will be limited only by photon statistics, as \T detectors are 
virtually noiseless \citep{{noah-phys-scripta98},{TOM}}. Effective exposure 
time of about 1000 sec per field in each of three \T telescopes will yield a 
resultant monochromatic UV magnitude of 25, or almost 15 magnitudes better 
than the $TD1$ survey and with much better spatial resolution. 

One of the major scientific goals of \T is to detect and 
measure variability on time scales ranging from $<1 \;\text{sec}-3$ years. 
The photometry capabilities (each detected photon is time-tagged with 
a precision of $\sim 128$ ms) and the possibility of repeated scans 
will enable the recording of UV light curves of variable sources; 
the time scale at which variability can be investigated will depend on 
the source latitude and brightness. In this article we are discussing 
the prospects of detecting the UV flares from active M dwarfs 
and possible collisions in extrasolar planetary systems by TAUVEX.  

\section{\T Observatory}
\label{sec:2}

\T Observatory is an array of three identical, 20 cm aperture, 
co-aligned F/8 Ritchey-Cr\'{e}tien telescopes mounted on a single 
bezel and enclosed in a common envelope. 
Each telescope consists of a primary mirror, a secondary mirror 
and 2 doublets of field-correcting lenses that maintain a relatively 
constant focal plane scale across the nominal field of view of 
$0.9^{\circ}$ and serve as Ly-$\alpha$ blockers when heated to $25^{\circ}$ C.  
The primary and secondary mirrors are both lightweighted zerodur 
coated with Al + MgF$_2$ with an effective reflectivity of better 
than 90\%. There are four filters per telescope offering a total of five 
different UV bands for observation (Fig.~\ref{fig:filters}). 

The detectors are photon-counting imaging devices, made of a CsTe 
photocathode deposited on the inner surface of the entrance window, a 
stack of three multi-channel plates (MCP), and a multi-electrode 
anode of the wedge-and-strip (WSZ) type. Properties of the
instrument are summarized in Table~\ref{table:instrument}.

Table~\ref{table:observatory} contains pre-launch predictions for the
in-orbit performance of \T\ns. The nominal mode of operation of \T 
is to scan along a single line of celestial latitude, with entire
sky coverage attained by rotating the mounting deck platform (MDP). 
Exposure time in a single scan is limited to 216 seconds at the 
celestial equator, but increases with declination to a theoretical 
maximum of $\sim 86,400$ seconds per day at celestial poles 
(see Table~{\ref{table:observatory}).

\begin{figure}[h!]
\begin{center}
\includegraphics[scale=0.3,angle=-90]{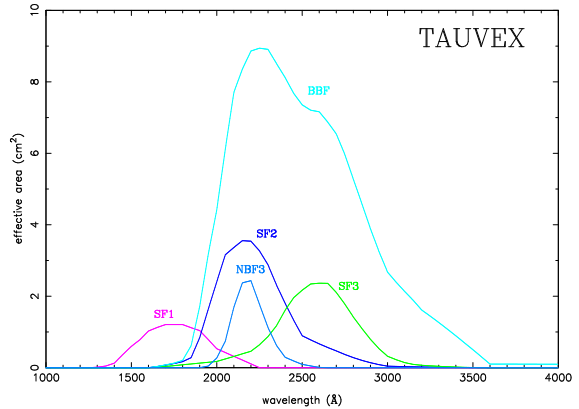}
\end{center}
\caption{The spectral range of the \T filters; effective areas at the beginning 
of the mission versus wavelength.}
\label{fig:filters}
\end{figure} 

\begin{table*}
\caption{Instrument Parameters.\label{table:instrument}}
\vskip 0.1in
\begin{tabular}{lll}
\tableline\tableline
Mirrors (2 per telescope)&:& hyperbolic, zerodur substrate, Al+MgF$_2$ coating\\
Optics &:& F/8 Ritchey-Chr\'{e}tian; 2 doublet CaF$_2$ lenses\\
Diameter &:& 20 cm aperture \\
Detector &:& 3-stack MCP with 25 mm CsTe photocathode\\
 &:&and Wedge-and-Strip anode\\
Observational mode & : & Scanning\\
Total payload weight &:& 69.5 Kg\\[0.02in]
\multicolumn{3}{c}{Locations of the filters}\\  
Telescope:&&Filters:\\[0.02in]
\hspace{1.cm}T1 & :&SF1; SF2; BBF\\
\hspace{1.cm}T2 & :&SF2; SF3; BBF\\
\hspace{1.cm}T3 & :&SF1; SF3; NBF3\\
\tableline\tableline
\end{tabular}
\end{table*}

\begin{table*}
\caption{Predicted Performance}
\label{table:observatory}
\vspace{0.1in}
\centering
\begin{tabular}{lll}
\hline\hline
\multicolumn{1}{l}{Wavelength coverage} &
\multicolumn{1}{l}{: $1250-3500$ \AA} \\
\multicolumn{1}{l}{Field of View (FOV)}&\multicolumn{1}{l}{:  $0.9^{\circ}$} \\
\multicolumn{1}{l}{Angular resolution}&\multicolumn{1}{l}{: $\sim 10''$}\\
\multicolumn{1}{l}{Expected source position localization}& \multicolumn{1}{l}{:  $\sim 5''$ (for a 
$15^m$ object)} \\
\multicolumn{1}{l}{Time resolution}&\multicolumn{1}{l}{: 128 milliseconds}\\
\multicolumn{1}{l}{Minimal exposure time$^{*}$}&\multicolumn{1}{l}{: 216 seconds}\\[0.02in]
\multicolumn{3}{c}{Filters central wavelengths and bandpasses}\\
Filter: & Wavelength (\AA)  &Width (\AA)\\[0.02in]
\hspace{0.9cm}NBF3&:\hspace{0.5cm} 2200 &:\hspace{0.5cm} 200 \\
\hspace{1.cm}SF1 &:\hspace{0.5cm} 1750 &:\hspace{0.5cm} 500 \\
\hspace{1.cm}SF2 &:\hspace{0.5cm} 2200  &:\hspace{0.5cm} 400 \\
\hspace{1.cm}SF3 &:\hspace{0.5cm} 2600  &:\hspace{0.5cm} 500 \\
\hspace{1.cm}BBF &:\hspace{0.5cm} 2300  &:\hspace{0.5cm} 1000 \\[0.02in]
Point source sensitivity (200 s)&\multicolumn{2}{l}{:  \hspace{0.4cm} (For an O type star)}\\
\hspace{2.95cm}BBF& \multicolumn{2}{l}{: \hspace{0.95cm}    $V \approx 22.2$}\\
\hspace{2.97cm}SF3& \multicolumn{2}{l}{:   \hspace{0.95cm}  $V \approx 20.2$}\\
\hspace{2.97cm}SF2& \multicolumn{2}{l}{:   \hspace{0.95cm}  $V \approx 20.8$}\\
\hspace{2.97cm}SF1& \multicolumn{2}{l}{:   \hspace{0.95cm}  $V \approx 20.3$}\\
\hspace{2.9cm}NBF3& \multicolumn{2}{l}{:   \hspace{0.95cm}  $V \approx 19.2$}\\
\multicolumn{1}{l}{Bright limit (point source, BBF) }&
\multicolumn{2}{l}{:\;\;$F_{\l}=1.3\times 10^{-11}$ erg cm$^{-2}$ s$^{-1}$ \AA$^{-1}$}\\  
\hline\hline
\end{tabular}\\
\vspace{0.1in}
*{this exposure time is for $\d =0^{\circ}$ for a source at the centre of FOV; 
it increases with declination and reaches theoretically $\sim 86000$ seconds on the Poles.}
\end{table*}
 
The primary data product of \T pipeline is a
calibrated photon list where each photon is tagged by sky position
and time, after correction of all instrumental effects. Tools will be
provided to extract multi-band images of the sky from these photon
lists. We will also create special tools
intended for analysis of variable sources. The first of these is
a real-time tool in which the brightness of any point source in the
field will be checked for variability both between frames, that is,
within an observation, and by comparison with predicted count rates
from historical observations. If a significant event, over and beyond
normal noise, is found, a responsible scientist will be immediately
notified and follow-up observations may be scheduled with other
ground and space based telescopes, such as, for example, Indian 
Astronomical Observatory (IAO) at Hanle. The other tool will run as 
part of the normal pipeline, perhaps days after the observation. This will
produce a time history of every source found during normal
observations, which can then later be examined for variability.

\subsection{\T Team Science Investigations}
\label{sec:4}

As an observatory class mission, \T has a broad scientific program. 
\T is different from most missions in that it is only a scanning 
instrument; pointed observations are not possible. Despite this 
major disadvantage, the location of \T in geostationary orbit 
allows for exceptionally deep observations of areas near the poles 
with very low backgrounds. Thus, one of the major scientific programs 
of \T is the Deep Exposure Polar Survey ({\it DEPS}), in which 
$\sim 1400$ square degrees around each Celestial Pole is intended to be covered 
with a uniform accumulated exposure of 5000 sec in the 
first year of operation, followed by a repeat of another 5000 sec in the 
second year. This is expected to achieve a depth of $25^m$ in three UV bands.  
(All magnitudes are quoted as per the monochromatic scale of Hayes \& Latham 
\cite{monochromatic}.) Another major survey is the Galactic Plane Survey 
({\it GAPS}), which will also be performed during the first two years of 
operations and will cover $\sim 1500$ square degrees with an average 
accumulated exposure of 1000 sec. It is expected that in total 5-7 years 
of \T lifetime the whole sky will be covered. Most of the surveys
will be performed in three \T principal filters, SF1, SF2 and SF3,  
which span the spectral region from somewhat longer than Lyman $\a$
to 320 nm with three well-defined bands. Three filters define two 
colour indices in the UV, and the combination of these measurements with 
data from the optical and infrared allows the derivation of even more 
colour indices.

\section{UV Emission from Celestial Collisions}

The phenomena of one celestial body colliding with another is quite 
common on all astronomical scales, from asteroids bombarding planets to
recently discovered collisions of massive galaxy clusters \cite{{bullet},
{cl0024}}. The origin of blue
stragglers in the globular clusters is due to merging of two, or even three,
MS stars of $\sim 0.8$ \msun. Perhaps, half the stars in the 
central regions of some globulars have undergone one or two collisions over a period
of $10^{10}$ years \citep{sivaram-ref3}. White dwarf (WD) binaries with periods
$< 5$ mins will merge in a few thousand years, and neutron star--white dwarf
(NS--WD) binaries with periods of $\sim 10$ mins have been observed 
\citep{sivaram-ref4}.  

Over the past few decades, a standard model for planetary system 
formation has emerged, in
which the final stage is now strongly associated with a significant 
number of giant impacts on each of the young forming planets \cite{stern}. 
Giant Impact Models (GIM) 
are often invoked for explaining the formation of planetary systems, the Mars-sized 
terrestrial impactor believed
to be responsible for the formation of the Moon was such an impact \cite{stevenson87}. 
However, even at later stages of planetary system life, such giant 
collisions are possible. In the past decade about 200 ESPs have 
been detected and they are a rapidly growing sample. Most of these ESPs
are giant `Hot Jupiters', locked in very short-period orbits with their star. 
It is reasonable to expect planetary and stellar-planetary collisions in such systems.  

Collisions on all scales can be very energetic. A recent example in our own 
Solar System is that
of the comet Shoemaker-Levy which collided with Jupiter. If a WD smashes into
a MS star like our Sun with an incoming velocity of, say, $\ge 700$ km/sec, the massive
shock wave would compress and heat the Sun, the nuclear reactions will speed up,
and, in about an hour, the Sun would release a thermonuclear energy of about $10^{49}$ ergs,
and the resulting instabilities would blow the Sun apart in few hours \citep{sivaram}. This
equivalent of a nuclear explosion of a star was recently revoked in a slightly different
context by Brassart and Luminet (2008) for the case of a star getting too close 
(or colliding) to a 
supermassive black hole whose tidal forces would create stresses inside the star that 
may trigger a nuclear explosion and give out a burst of an
electromagnetic radiation. Gamma-ray bursts (GRBs) once were thought
to be of a Galactic origin; one popular model being of an asteroid-size
objects (with $\sim 10^{32}$ gm of mass) impacting a neutron star (NS). 
In such case, with the impact velocity
at a surface of a NS of $\sim 0.1 \,c$, the released kinetic energy would be
$\sim 10^{52}$ ergs (or $T\sim 10^9$ K), with most of the radiation in the form of
a few hundred Kev $\g$-rays. Recently, the impact theory was revived by proposing 
an alternative model for the short-duration GRBs, explaining them as supernovae 
followed the collision of compact objects \citep{sivaram}.

\subsection{Planet-planet collisions}

Stern (1994) has estimated that for planetary impacts with total collisional energy above 
$5\times 10^{34}$ ergs the dominant heat sink would be radiation to space. We will 
consider here a Jovian planet (with mass $\sim M_{\rm J}$) being hit by a 
terrestrial-size planet (with mass $\sim m_{\rm E}$), as in such a case 
the total impact energy is of the order $E_{\rm imp}
\sim GM_{\rm J}m_{\rm E}/R_{\rm J} \sim 10^{40}$
ergs. Giant impacts range from glancing events to direct head-on collisions. 
It is reasonable to assume hyper-velocity impacts, as, for example, in the 
end-member archetypes of planetary collisions. However, even the proposed 
origin of the Moon requires an impact velocity just
barely above the two-body escape velocity \citep{nature-paper}. Still, 
even if a giant impactor approaches the target with negligible velocity, 
its decent into the target's \gv potential well will cause it to impact at
near escape velocity. Therefore, in our calculations
we assume the starting impact velocity as two-body mutual escape velocity, 
which for terrestrial and Jovian-like bodies will be
\be
v^2_{\rm imp}\approx v_{\infty}^2+v_{\rm esc}^2\,,
\label{eq:1}
\end{equation}
where $v_{\infty}$ is the random velocity of the encountering bodies. 
Assuming $v_{\infty}=0$, we obtain
\be
v_{\rm esc}=\sqrt{\fr{2G(M_{\rm J}+m_{\rm E})}{(R_{\rm J}+r_{\rm E})}}\approx 40 \,
\text{km/sec}\,.
\label{eq:2}
\end{equation}

The friction between the impactor and the atmosphere of the impacted planet, which
increases in density as the target descends, generates an intense shock wave, heating
the target. It was assumed in Zhang \& Sigurdsson (2003) that since the energy deposit 
during the impactor descent is much greater than the corresponding Eddington 
luminosity of the target, $L^{\rm J}_{\rm Edd}$, the peak luminosity resulting 
from the hot spot created by the impactor will be 
$L_{\rm pk}\sim\eta L^{\rm J}_{\rm Edd} \approx L^{\rm J}_{\rm Edd}$. 
This luminosity will be produced by the hot 
spot with a peak temperature of $\sim 10^5$ K and in a time-scale
of few seconds, giving rise to an electromagnetic flash at far-UV 
$\l \sim 450$ \AA. However, it is quite likely 
that $\eta$, the factor correcting for radiative inefficiencies in 
Zhang \& Sigurdsson (2003), is not unity. Besides, the gas opacity of 
the atmosphere is crucial for the escape of radiation from the shocked matter, 
which was not considered in Zhang \& Sigurdsson (2003). 

Let us assume the atmospheric entry at velocity $v$ and angle $\phi$ 
(see Fig.~\ref{fig:entry} in Appendix~A). The force of drag on an impactor is 
$F_{\rm D}=1/2 C_{\rm D} \rho_a A v^2$, where $\rho_a$ is the atmospheric 
density and $C_{\rm D}$ is the coefficient of atmospheric drag 
($C_{\rm D}$ ranges from $1/2$ for spheres to $\approx 1/7$ 
\citep{chyba} for cylinders). The power released due to the drag will be 
\be
\dot E = \half C_{\rm D}\rho_a A v^3\,,
\label{eq:3}
\end{equation}
where $A$ is the front area of the impactor. If we assume that all dissipated 
energy is lost as radiation, $\eta\rho_a A v^3\sim \s A T_{\rm R}^4$, we obtain
\be
T_{\rm R} \sim \left(\fr{\eta\rho_a v^3}{\s}\right)^{1/4}\,,
\label{eq:4}
\end{equation}
where $\s$ is the Stephan-Boltzmann constant and $\eta$ is a fraction of energy going into
heat. We assume $\eta \approx 1$ ($T_{\rm R}$ it is relatively insensitive to it). 
The temperature estimates for different impact velocities, taken in the range
$10-40\,\text{km/s}$, are given in Table~\ref{tab:temperatures}. 
We see from the table that $T_{\rm R}$ strongly depends
on the velocity of the impactor and a wide range of temperatures is possible.

\begin{table}
\begin{center}
\caption{This table presents estimates of $T_{\rm R}(K)$ from Eq.~\ref{eq:4} as a
function of density $\rho$ (g/cm$^3$) and impact velocity $v$.} 
\begin{tabular}{l|l|l}
$\rho \setminus v$& 10 km/sec & 40 km/sec\\
\hline 
$10^{-8}$  & $\sim 3,600$ & $\sim 10,000$\\
$10^{-6}$  & $\sim 11,500$ & $\sim 32,500$\\
$10^{-4}$  & $\sim 36,000$ & $\sim 100,000$\\
$10^{-3}$  & $\sim 65,000$ & $\sim 180,000$\\
\hline
\end{tabular}
\label{tab:temperatures}
\end{center}
\end{table} 

For a thermal spectrum, a region of $2-6\times 10^4$ K represents the 
near-UV part of the spectrum. With 
$T_{\rm R}\sim 10^5$, the predominant radiation will be in the far-UV region, 
however, the prompt radiation, firstly, strongly depends on initial velocity, 
and, secondly, as was shown in detail in Chevalier \& Sarazin (1994), since the gas 
is optically thin in the top layers of atmosphere, the initial radiation 
will be concentrated in the form of emission lines. At $\rho \approx 10^{-6}$, 
merely a $\sim$ 
second after the start of a descent, the shocked atmosphere is optically thick, 
shock becomes a source of black-body continuum radiation, and the favoured emission 
is in $1000-3000$ \AA\, initially. This radiation can escape in the 
prompt phase from the vicinity of the shock front, but radiation bluer than 
$\sim 900$ \AA \, is trapped very close to the shock and is absorbed in a 
narrow pre-shock region (thus only the longer wavelength UV radiation 
can escape). At this prompt phase, the luminosity can reach 
$L_{\rm UV} \sim 10^{31}$ erg/sec, and the total integrated energy released 
during descent may reach $\sim 10^{32}$ ergs (see calculations 
App.~A). The flux of UV photons from the Sun at the Earth (at 1 A.U.) 
is $\sim 3\times 10^{14}$ phot/cm$^{2}$ s, from a distance of 10 pc it 
would be $\sim 50$ phot/cm$^2$ s. The flare flux at Earth in the UV 
at this distance would be $\sim 10^3$ phot/cm$^2$ s, thus outshining 
any MS host star later than B5. The maximum expected flux at Earth 
is in a range:

\begin{center}
\begin{table}[h!]
\small
\begin{tabular}{c|c|c}
          	  & at 10 pc                   & at 1Kpc   \\
\hline
Host star & $1-50$ ph/cm$^2$ s & $10^{-2}-10^{-3}$ ph/cm$^2$ s\\
(G, K) & &  \\
Coll. event & upto $10^3$ ph/cm$^2$ s & upto $1$ ph/cm$^2$ s  \\
at 3000 \AA&&\\
\hline
\end{tabular}
\end{table}
\end{center} 

\vskip 0.1in

The radiation in the prompt phase of the collision can be defined as 
a flare--a sudden, rapid and intense variation in brightness. Flares 
are a one-time event. The flares from the collisions will be followed 
by repeated flashes that can be produced through two different mechanisms. 
One type of flash is caused by the target planet's rotation. Since the 
typical spin period of giant planets is $\tau_2 < 1$ day, one expects a 
periodic luminosity modulation of the UV light curve, caused by an 
expanding hot spot entering and leaving the line of sight over a time-scale 
of $\tau_2$. This may also be detectable in optical bands. 

The second type of flash is caused by a fall-back of material in the 
case of the impactor reaching the surface of the target. Prompt
flare radiation pressure would drive a fraction of the impacted material 
outwards; as the radiation pressure drops, the material will fall back 
on the surface at $\sim v_{\rm esc}/2$. This fall-back can lead
to reheating and repeated flash(es) of decreasing brightness. If we 
consider a universal hydrodynamic scaling relation, the distance traveled 
in time $t$ is

\begin{equation}
R=\left(\fr{1}{\a_s}\fr{E}{\rho_a}\right)^{1/5} t^{2/5}\,,
\end{equation}
where $\a_s$ is $1.3-1.7$ for a spherical blast wave. The free-fall 
time would be of order of $\tau_3\sim$ few hours. The first flash 
will have the energy of $\sim 1/4$ of the flare.
    
Thus, we establish three important time-scales, 

\begin{center}
\begin{tabular}{l|l|c}
\hline
$\tau_1 $& flare&    upto $10^3$ s\\
$\tau_2$ & modulation due to spin &    $ \lesssim  1$ day\\
$\tau_3$ &  flashes due to fall-back &  $\sim$ few hours\\
\hline
\end{tabular}
\end{center}

\vskip 0.1in

More exotic cases of planetary collisions are possible. The discovery 
of several ``very hot Jupiters" \citep{chtonian} brings an interesting
possibility that because of their proximity to the host star, they may have their 
atmospheres evaporated; they are the new class of planets called ``chthonians". 
In this case, the direct impact onto a hot chthonian surface will produce a flare 
with a temperature $T\approx 1500-2000$ K, with consequent flashes of decreasing
brightness. 

If two `Hot Jupiters', tidally locked
with the host star, collide, such collision could generate $\sim 10^{45}$ ergs 
with peak temperature at $\sim 2\times 10^5$. For an event duration of 10 hours, 
we can expect an EUV fluence of $10^5$ phot/cm$^2$ from a collision at 
10 Kpc, while
the fluence from the host star will constitute most probably only $\sim$1 phot/cm$^2$. 
For the \T sensitivity range this will correspond to about $10^4$ phot/cm$^2$. 
If the rotation 
period is about one day, this flux will be modulated over a day.

\subsection{Stellar-planetary collisions}

Many extra-solar Jovian planets are in very tight orbits around the star. 
Out of the total of 246 extra-solar planetary systems discovered to date, there are,
at least, 30 planets with orbits less than 1 A.U. from the host star. Assuming the  
velocities are of order $\sim 200-300$ km/sec, the planet can spiral down in 
$\sim 10^8-10^9$ years (due to the gravitational radiation). It is conceivable 
that a collision between the planet and the star might be frequent, where 
the collisional time-scale will be of the order $10^6$ sec. 
For the impact velocity of
$\sim 600$ km/sec, the kinetic energy release in the collision will be $\sim
10^{46}$ ergs with a peak temperature, following Eqs.~(\ref{eq:3}) and (\ref{eq:4}),
of $T_{\rm peak} \sim 6\times 10^5$ K. This would imply the soft X-ray flux  
of $\sim 10^4$ ergs/cm$^2$/sec at a distance of 10 pc. The flux at Earth 
from the distance of 10 Kpc is $\sim 10^{-6}$ ergs/cm$^{2}$/sec. This soft X-ray flash
can last a few hours and is $10^8$ times more intense than the strongest solar flare. 
The initial soft X-ray
flash will be followed by intense EUV-UV radiation for few days, with flux 
at Earth of the order $\sim 0.3$ ergs/cm$^{2}$/sec at 10 pc distance. 
From 10 Kpc, the UV flux will be $\sim 3\times 10^{-7}$ ergs/cm$^{2}$/sec. 
In the \T bands we can expect $\sim 10^8$ phot/cm$^{2}$ sec from a distance of 
10 pc and $10^2$ phot/cm$^{2}$ sec from a distance of 10 Kpc. We illustrate
these values in Table~\ref{table:distances}.     

\begin{table*}
\caption{Expected flux at Earth in different bands
\label{table:distances}}
\vskip 0.1in
\begin{tabular}{c|c|c|c}
    Distance	  & X-ray     & EUV  & TAUVEX \\
\tableline
10 pc & $80-100$ ergs/cm$^2$ s & $0.3$ ergs/cm$^2$ s & $10^8$ phot/cm$^2$ s\\
10 Kpc &$\sim 10^{-5}$ ergs/cm$^2$ s & $3\times 10^{-7}$ ergs/cm$^2$ s & $10^2$ phot/cm$^2$ s \\
\tableline
\end{tabular}
\end{table*}
 
Impacts of comets with young stars having dust debris disks are also energetic collision
events. In such case, however, the diffuse structure of a comet with its extensive coma
would reduce the intensity of the power, or flux, of the flare that may result from a 
collision. The volatile elements would vaporise, or sublimate, on approaching the star (a
comet could even completely disintegrate). Use of the equations in the Apendix~A shows that
the dependence on the area (and mass) would give a significantly lower temperature (and total 
released energy) than in the case of an impact of a larger planetary body. Most of the 
radiation would be in the infrared or longer optical wavelengths. (For example, collision 
of a Shoemaker-Levy comet with Jupiter resulted in intense IR radiation.) There could be 
also an initial optical flash, but the UV flux would be substantially smaller.

\section{M-dwarf flares}
\label{subsec:dMflares}

Another type of short-scale transients are stellar flares from late-type stars, for example, 
M dwarfs. M dwarfs account for more than 75\% of the stellar population in the solar
neighbourhood (up to 1 Kpc). These stars are known to possess strong magnetic fields with
high coronal activity and associated UV line emission \citep{Mitra-Kraev} and a vast
majority of UV short transients is associated with a single physical type of astrophysical 
source, that is, stellar flare eruptions on K and M stars. These flares exceed the quiescent
phase luminosity by two or three orders of magnitude and are a thousand times more 
energetic than solar flares. The peak temperature (in the active or disturbed regions) 
of the flare reaches $10^4-3\times 10^4$ K, which implies an UV peak, at luminosities
of $10^{25}-10^{26}$~W. For such a flare on, for example, Proxima Cen, this could 
correspond to a 
flux at Earth of $10^5-10^6$ UV phots/cm$^2$ for a period of several hours. The M 
dwarf flare 100 pc away will produce UV flux at Earth of $10-10^2$ phots/cm$^2$, 
which is  detectable by TAUVEX.      
Studying the frequency of dMe flares can be important from several standpoints. 
\begin{itemize}
\item The level of chromospheric activity may vary with dMe stellar age and 
mass \citep{Gizis}. Observations (see Welsh et al. (2007) and references within) indicate
that activity strength (as measured by the ratio of H$\a$
luminosity to the stellar bolometric luminosity) increases with the spectral 
class---earlier dMes have higher flare energies. However, as was discussed in 
Kulkarni \& Rau (2006), there may be a fraction of late-type dwarfs that retain their 
activity for a longer period or may have had activity induced later in 
their lives.

\item A derivation of the UV flare frequency rate on M dwarfs is also important for the 
study of habitability zones on possible associated ESP systems \citep{Turnbull}.

\item Lastly, the estimated dMe transients annual rate \citep{DLS}
of $R_{R} \sim 10^{8}\,\text{yr}^{-1}$ implies the existence of immense fog 
of M dwarfs that hinders any extragalactic transient search---the foreground 
fog of dMe flares ensures that the false positives outnumber genuine 
extragalactic events by at least two orders of magnitude. The clear identification 
of all suspect dwarfs, especially over the extragalactic fields,
will ensure that the extragalactic surveys, like, for ex. PANSTARRS or LSST, will not
drown in false detections. 
\end{itemize}  

The recent detection of a flare star by GALEX \citep{{galex-uvceti-flash}} 
has shown the power of UV observations in such studies. The peak of the 
radiation is at $\l < 4000$~\AA \,\citep{Oord} and optical photometry will 
only sample a wing of this distribution against the bright stellar photosphere. 
By contrast, the continuum and line emission during a flare in the UV will 
cause a much greater spike in the intensity, by a factor of 1000 or more,
which will be seen against a much fainter stellar photospheric background.

\section{Prospects of Observations with TAUVEX}

\subsection{Estimates of frequency and detectability of planetary collisions}

A large number of collisions should be happening in dynamically young systems (like
planetary systems in formation), but there could be collisions in mature systems
as well. For example, $\tau$ Ceti has more than 10 times the amount of dust as the Sun.
Any planet around $\tau$ Ceti would suffer from large impact events roughly 10 times more
frequently than Earth. Even more interesting is $\e$ Eri, which has $\sim 10^3$
more dust than is present in the inner system around our Sun. It has a detected 
planetary system as well; with one $1.55 M_{\rm J}$ mass planet at $3.4$ A.U.   
At, say, 3000 \AA, the flux from $\tau$ Ceti at Earth is $3.4\times 10^{-11}$ ergs/cm$^2$ sec,
which is equivalent to $\sim 6$ phots/cm$^2$ sec. The UV flare in its system may produce 
up to $\sim  4\times 10^{-8}$ ergs/cm$^2$sec, or up to $\sim 6\times 10^3$ phots/cm$^2$sec
in the UV. The flux contrast is dramatic. Most known exoplanets orbit stars 
which are roughly similar to our own Sun, that is, MS stars of spectral 
categories F, G, or K, with the current count (year 2007) of ESP candidate 
systems of 252, most of them within 1 Kpc distance from the Sun. 

The flux at Earth from a self-luminous thermally dominated object is
\begin{equation}
F_{\l}=\left( \fr{R}{D}\right)^2\int_{\l_1}^{\l_2} B_{\l}(T) d\l\,,
\end{equation}
where $D$ is the distance to Earth, $R$ is the radius of the source, 
$B_{\l}(T)$ is the Planck function, and the integration limit refers 
to the width of a given bandpass. Fig.~\ref{fig:bbody} depicts the 
UV flares flux curves from a terrestrial-size hot spot with $T=20000$ K 
as seen at different distances from Earth. It also shows the black 
body flux curves for a solar type (G8V, $\tau$ Ceti 
taken as an example) star, a cooler star (K2V, $\e$ Eri as an example) as seen 
from the same distances, along with UV sensitivity of \T in all filters for 
1 Ksec exposure. For comparison, Fig.~\ref{fig:limits} shows UV sensitivity
of TAUVEX in all filters (and filters wavelength coverage) for different exposure times. 
It is clearly seen that for an average exposure of 1000 sec the flares are easily 
detectable by \T\ns, even if they occur at 1 Kpc distance. These calculations 
demonstrate that hot spots resulting from the terrestrial-size object impacting 
the giant planet are sufficiently bright to be detected from a flux sensitivity 
standpoint.
 
\begin{figure*}
\begin{center}
\includegraphics[scale=0.4,angle=-90]{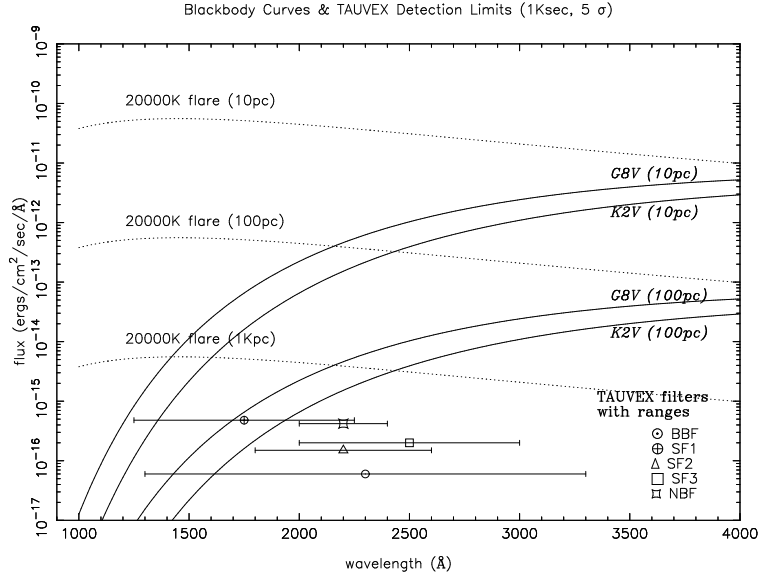}
\caption{Thermal flux curves for G8V and K2V stars and UV collision flare 
($2 \times 10^4$ K) as seen at different distances from Earth. Also shown are 
the sensitivity limits for a 5 $\s$ detection by \T in all filter bands for 
1000 sec exposure (the span of each line shows the wavelength coverage of that
filter). This shows a clear detectability of the flares from collisions
in the situation of an average non-negligible sky background.}  
\label{fig:bbody}
\end{center}
\end{figure*} 
 
\begin{figure*}
\begin{center}
\includegraphics[scale=0.45,angle=-90]{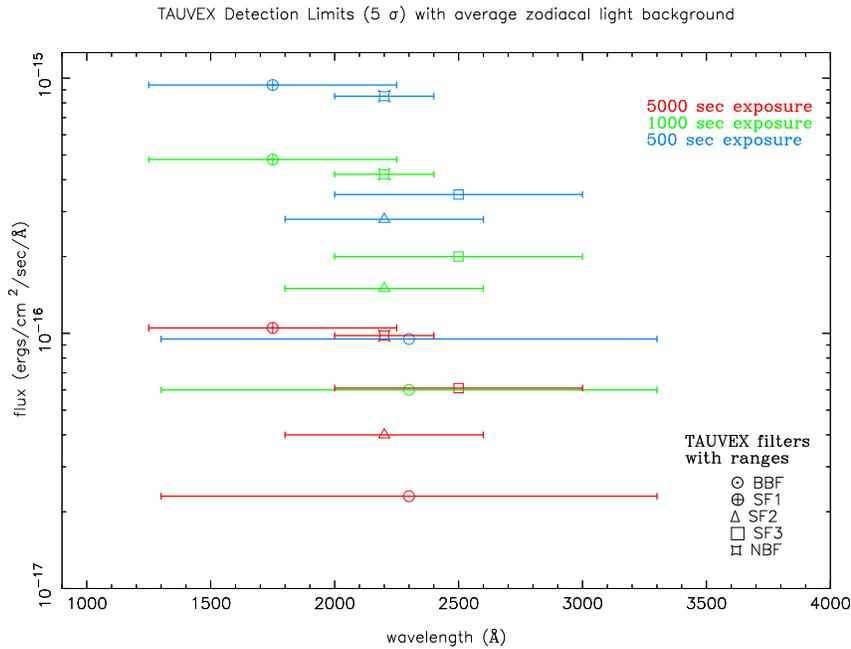}
\caption{Sensitivity limits for a 5 $\s$ detection in all filter bands for proposed 
\T surveys. The span of each line shows the wavelength range of that filter.}
\label{fig:limits}
\end{center}
\end{figure*}

In young planetary systems the central star is surrounded by planetesimals 
with a few embryos of mass $10-100 M_{\oplus}$. We can easily assume existence 
of $10^8-10^9$ planetesimals in $\sim 100$ km diameter, or more! In a 
$(1 \text{A.U.})^3$ volume the number is $n=10^{-32}$ cm$^{-3}$. The mean 
time between the collisions is of order $t_c\sim1/(n\sigma v)$. With 
$v=10^2$ km/sec, $\s \sim (GM/v^2)^2 \sim 10^{16}-10^7$ cm$^2$, and time 
$\sim 10^6$ sec, we can have 10 colls/year (of $\sim 10^2$ km size objects) 
in a typical system. If there are $10^7-10^8$ such systems in a whole Galaxy, 
we can expect $10^8-10^9$ such events per year. 

Using the annual rate of events, the detection rate can be found. 
Let $\tau$ be the exposure time per FOV (say, $1000$ sec). The number of 
flares per exposure is  

\begin{equation}
N_{f}=\fr{R_{f}\,\tau \Delta\Omega}{4\pi}\,,
\label{eq:num_exp_flashes}
\end{equation}
where $\Delta\Omega\approx 2\times 10^{-4}$ sr is the \T FOV, and $R_{f}
\sim 10^8$ yr$^{-1}$ is an annual rate, estimated above. In one year 
of observations we can expect about $90-350$ such events. We can differentiate 
these flares from other transients (like, for instance, flares from M dwarfs, 
discussed in the next section) because of the temporal modulation due to 
the planet's rotation, or even the orbital period, if it is a hot Jupiter. 
Occurence of repeated flashes (flares) of diminishing intensity due to the 
infall (and subsequent rebound) of debris falling back on the impactor is 
an additional clue. Every such collision will be followed by a long-duration 
infra-red emission, which could perhaps be detected on the ground as a follow-up.

\subsection{M-dwarf flares}

During TAUVEX surveys, a simultaneous monitoring of a $\sim 1^{\circ}$ field 
of view in three UV bands will allow the accurate determination of UV fluxes 
and colours at high time resolution ($\sim 0.1$ sec) of any detected flare. 
Three UV bands mean two UV colours, and obtaining flare colours allows 
estimates of temperatures and sizes of active regions. For example, 
in assumption of the black body emission (which is only one of several possible flare
model scenarios, but still being widely used in the studies of flares, see 
Robinson et al. (2005) and references within), we can determine the relation 
between the black body and the measured SF1/SF2 and SF2/SF3 flux ratios. 
This is calculated by convolving the black body spectrum for different temperatures 
with the effective area curves for three TAUVEX principal survey filters, SF1, 
SF2 and SF3, and then normalizing to the total effective area of each filter.
This gives the expected flux for a black body in each filter, which can later be 
directly related to the average flux determined from the measured count rates. 
The relation between temperature and flux ratios is shown in Fig.~\ref{fig:fluxratio}.

\begin{figure}[h!]
\begin{center}
\includegraphics[scale=0.33,angle=-90]{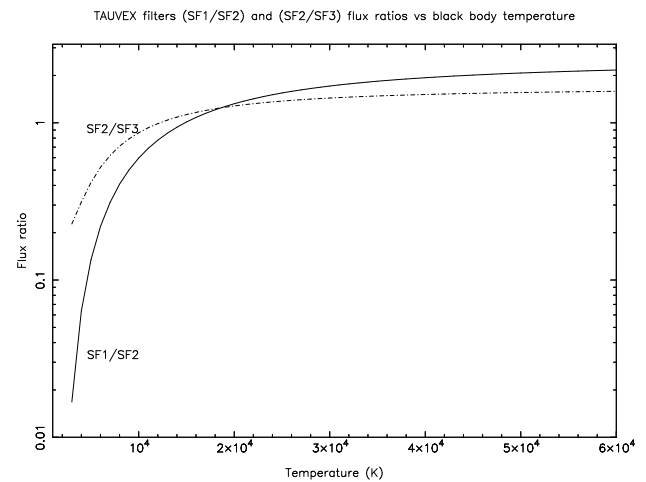}
\caption{Relation between black body temperature and the SF1/SF2 and SF2/SF3 flux    
ratios, determined by convolving the appropriate blackbody distribution
with TAUVEX principal survey filters (SF1,SF2,SF3) effective area curves.}
\label{fig:fluxratio}
\end{center}
\end{figure} 

The comparison of measured flux ratios of observed flares with the theoretical will
give an estimate of the effective temperature and a black body of a given temperature 
has a well-defined emission. Having determined the temperature, we can define the 
effective stellar surface coverage required to produce the observed flux at Earth, 
assuming (or determining) the distance to the star. As a result, we can then 
determine the total energy of the flare by integrating over the approriate black 
body curve.   

We can estimate the flares detection rate in the \T surveys. If we take the exposure
time per FOV as $\sim 5000$ sec in {\it DEPS} and $\sim 600$ sec in {\it GAPS}, 
the number of flares per exposure is (see Eq.~\ref{eq:num_exp_flashes})
\begin{equation}
N_{F}=\fr{R_{F}\,\tau \Delta\Omega}{4\pi}\,,
\label{eq:num_exp_dM}
\end{equation}
where $R_{F} \sim 10^8$ yr$^{-1}$ is an annual all-sky rate of fast transients 
in B band, estimated elsewhere \citep{Becker,DLS}. 
Then $N_{F}^{DEPS}=0.25$ and $N_{F}^{GAPS}=0.03$. Assuming 20\% efficiency in 
utilizing time in {\it DEPS} and 30\% in {\it GAPS}, we
can estimate that in a year of observation we will observe $N^{DEPS}_{F}\sim
950$ and $N^{GAPS}_{F}\sim 290$. 

Identification of any dMe flare in {\it DEPS} (with its high ecliptic latitude) 
may help extragalactic surveys, such as, for example, DLS, or even, GRB searches, 
as some of dMe flares were already mistaken for GRBs \citep{DLS}. The two colours
obtained by \T may be used to discriminate against genuine extragalactic 
transients---the Rayleigh-Jeans tail of a blackbody, for example, has a distinctly 
different spectrum ($f_{\nu}\propto\nu^2$) than, say, synchrotron radiation 
($f_{\nu}\propto \nu^{\a}$, with $\a\sim -1$).

\section{Conclusion}

In this paper we considered two sources of the UV short-scale 
transient events, namely, the UV flares from the active late-type stars 
and from possible collisions in extrasolar planetary systems. UV transients 
can be broadly divided into three categories: `regular', `expected' and `unexpected'
events. `Regular' events would include CVs, recurring novae, etc. These observations are
part of a regular TAUVEX observational plan; they will be scheduled according to the submitted
proposals. The scientific output of these is the responsibility of the respective scientists 
who submit the proposal. `Unexpected' events include SNs, GRBs, UV/X-ray flares resulting from
the tidal disruption of a star passing too close to a supermassive black hole (the latter
event can result in a series of UV/X-ray flares, initially from a star being blown apart by an
internal nuclear explosion \citep{luminet}, subsequently from the stellar debris 
falling back into the black hole \citep{GALEX-BH}, and finally from the debris
stream collision with itself long after the disruption \citep{kim}). These transients 
are impossible to predict, they are of a long duration (from days to years), and 
some, for example, SNd Ia and GRBs are intrinsically faint in the UV.  

The category of
the `expected' UV transients, that would include the flares from the late-type stars, both
known and unknown, and UV flashes from the collisions in the extrasolar planetary
systems, is however the one that is the subject of the present investigation. The latter
can include the planet-planet collisions, planet-star collisions and 
smaller objects/debris collisions (comets, asteroids, etc.). We have concentrated 
on the first two types
of collisions as these would produce the strong signal in the appropriate energy range,
detectable by TAUVEX. We discussed the dynamics
of planetary collisions in some detail, especially the collisions of terrestrial-mass
planets with Jovian planets and hot Jupiters. Dynamics of atmospheric drag, shock-wave
bounce, infall of debris, etc. are taken into account. The duration of the events as
well the UV fluence expected from events at 10 pc and 10 Kpc are estimated. The
flux from the UV flare event in contrast to emission from the host star is estimated.
Expected annual rate of such events is calculated for the FOV of \T detector. 
 
\T observatory will monitor large areas of the sky with different cadences and is 
well suited for variability studies on scales ranging from less than a second to months.
We will implement programs to monitor \T data in real-time and to notify the 
responsible scientist in the case of a flare. Follow-up observations with ground-based 
telescopes will be immediately initiated.

The next step in our preparations for \T observations is going to be the simulation
of the late-type star flare spectrum through available software 
(for example, the CHIANTI package) over the \T wavelength range. This range 
contains many emission lines together with the
continuum radiation. Assessing the relative contributions from continuum
and/or emission lines at near UV wavelengths is still an
open problem for flare research (see Robinson et al. (2005) and discussion therein). 
Convolution of the simulated spectrum with \T filters response 
functions will give us the emission lines contained within the
filters. We will derive a set of templates of \T filters ratios for different values
of electon pressures and different differential emission measures (DEM). 
Thus, if a flare is observed by \T\ns, the comparison of the theoretical flux ratios
with the measured ones can help in determining the value of the plasma electron density
during the flare event, in the assumption that the increase in electron pressure
during the flare has a major influence on the ratios \citep{galex-spectra}. 
In addition, since GALEX has produced a catalog of GALEX transient events 
\citep{galex-catalog}, the predictions for the stellar-planetary collision 
UV flashes could be tested against the existing GALEX data set, and this topic 
will also be addressed in the forthcoming paper. 

\section{Additional Information}

The most extensive description of proposed TAUVEX science, predicted performance 
and planned calibrations till date is presented in the special issue of the 
Bulletin of Indian Astronomical Society \cite{BASI-tauvex}.
 
Proposed \T launch date is January-February 2009. This date can be revised but the launch
is scheduled to be no later than April 2009. The latest updates on the launch date and 
technical information about the initial performance results and information about observing with 
\T can be found in the {\it TAUVEX Observer's Manual} \citep{TOM} 
and on the TAUVEX web site at: http://tauvex.iiap.res.in. 
Guest Investigator questions can be directed to \T help desk at 
tauvex@iiap.res.in, or to Dr.~M.~Safonova at rita@iiap.res.in.

\section{Acknowledgments}

It is a pleasure to thank many dedicated people who are working hard to make \T mission
happen, most of all, the \T Core Group, the GSAT-4 group at the ISRO Satellite Centre (ISAC)
and the engineers at El-Op. We also thank the anonymous referee for exremely helpful 
and useful comments, and especially for attracting our attention to a case of stellar
disruptions by supermassive black holes, which is now included in one of the TAUVEX
Key Areas observational plan.

\appendix
\section{Appendix. Detailed Calculations for Planetary Collisions}\label{app:appA}

\begin{center}
\begin{figure*}[h!]
\includegraphics[scale=0.6]{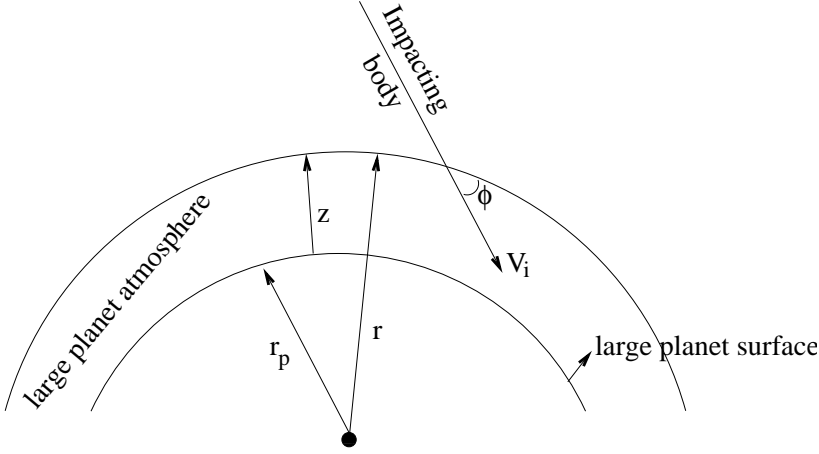}
\caption{Atmospheric entry at an angle $\phi$, $90^{\circ}$ is grazing, 
$0^{\circ}$ is head-on.}
\label{fig:entry}
\end{figure*} 
\end{center}
 
The above figure depicts the atmospheric entry of a body mass $m$ at an angle $\phi$
with initial velocity $v_i$ into the atmosphere of a larger body. Here 
$z$--height of the atmosphere, $r_{\rm p}$ is the radius of a target, $r$--radial 
coordinate. The most general equation of motion for the descent is
\begin{equation}
\fr{v}{\cos{\phi}}\dder{\phi}{t}=g-\fr{v^2}{r}-\fr{C_{\rm D}A}
{m_{\rm E}}\fr{\rho_a}{2}\fr{v^2}{\cos{\phi}}\,,
\label{eq:a1}
\end{equation}
where $\rho_a$ is atmospheric density, $A$ is interacting area of the impactor, 
$g$ is acceleration due to gravity and $C_{\rm D}$ is the coefficient of atmospheric drag. 
Assuming isothermal atmosphere, $\rho(z)=\rho_0 \me^{-\mu z}$ with $1/\mu=
R_g T/M_g$ the scale height, where $R_g$ is the gas constant, $M_g$ is the molecular
weight of the atmosphere and $T$--the temperature. After introducing the 
dimensionless Chapman parameters 
\be
\bar{v}=\fr{v\cos{\phi}}{v_i}\quad \text{and}\quad Ch=\fr{\rho_0}{2\mu}\fr{C_{\rm D}A}
{m}\sqrt{\mu r}\me^{-\mu z}\,,
\label{eq:a2}
\end{equation}
the equation becomes
\be
\bar{v}^2+\left(Ch\right)''+\bar{v}\left(Ch\right)'= \fr{1-\bar{v}^2}{v\bar{v}^2 Ch}
\cos^4{\phi}-\sqrt{\mu r} \fr{L}{D}\cos^3{\phi}\,,
\label{eq:a3}
\end{equation}
where $L/D\sim 1$ is the lift-to-drag ratio. The deceleration will be
\be
\dder{\bar{v}}{t}=-\sqrt{g\mu} \bar{v}^2\fr{Ch}{\cos{\phi}}\,.
\label{eq:a4}
\end{equation}
The time of flight is 
\be
t_2-t_1=\fr{1}{\sqrt{g\mu}}\int_{\bar{v}_1}^{\bar{v}_2}\fr{\cos^2{\phi}}{\bar{v}^2 Ch}
d\bar{v}\,
\label{eq:a5}
\end{equation}
and the distance moved for deceleration between $\bar{v}_1$ and $\bar{v}_2$ is
\be
\fr{x_2-x_1}{r}=\fr{1}{\sqrt{\mu r}}\int_{\bar{v}_1}^{\bar{v}_2}\fr{\cos{\phi}}{\bar{v}
Ch}d\bar{v}\,.
\label{eq:a6}
\end{equation}
The solution for deceleration is thus
\be
-\dder{v}{t}=-\mu v_i^2\sin{\phi}\left(\fr{v}{v_i}\right)^2 \log{\left(\fr{v}{v_i}\right)}\,.
\label{eq:a7}
\end{equation}
The $\log$ term here comes from several bouncing collisions with atmospheric molecules
that introduce $dv \propto v$ type of effect. This is valid for large Knudsen number,
when $mf_p/\text{particle size} >>1$), which is generally valid for the upper atmospheres. 
The maximum deceleration
\be
-\left(\fr{dv}{dt}\right)_{\rm max} =\fr{v_i^2\sin{\phi}}{2\me}
\label{eq:a8}
\end{equation}
occurs at $v/v_i=1/\sqrt{\me}=0.607$, that is when the velocity becomes $\sim 0.6 v_i$.
The total deceleration time can be found from
\be
t-t_i=\fr{1}{\mu v_i\sin{\phi}}\left[ \mathrm{E}_i\left(\fr{C_{\rm D}A}{m\sin{\phi}}
\fr{\rho_0}{2\mu}\g\right)-\mathrm{E}_i\left(\fr{C_{\rm D}A}{m\sin{\phi}}
\fr{\rho_0}{2\mu}\g_i\right)\right]\,,
\label{eq:a9}
\end{equation}
where 
\be
\mathrm{E}_i(\theta)=\int_0^{\theta} \me^{-y}/y dy\quad\text{and}\quad
\g=\me^{-\mu z}\,.
\label{eq:a10}
\end{equation}
We can find the terminal velocity (when it becomes constant) as
\be
v_t=\left(\fr{C_{\rm D}A}{m}\fr{\rho_0}{2g}\g\right)^{-1/2}\,.
\label{eq:a11}
\end{equation}
For example, for $m \sim 10^{17}$ kg and $A\sim 10^5$ km$^2$, 
the $v_t \sim 10$ km/sec.

Same calculations can be done for the grazing impact, where $\cos{\phi}\approx 1$. 
Assuming $d\phi/dt \approx 0$,
\be
\fr{v}{v_0}=\left[1+\mu r \fr{L}{D} \fr{C_{\rm D}A}{m}\fr{\rho_0\g}{2\mu}\right]^{-1/2}\,
v_0=\sqrt{g r_p}\,,
\label{eq:a12}
\end{equation}
the deceleration is
\be
-\dder{v}{t}=g\left(\fr{L}{D} +\left[\mu r \fr{C_{\rm D}A}{m}\fr{\rho_0\g}{2\mu}
\right]^{-1}\right)^{-1}\,
\label{eq:a13}
\end{equation}
and 
\be
\left(-\dder{v}{t}\right)_{\rm max}=\fr{L}{D}\,.
\label{eq:a14}
\end{equation}
Here maximum deceleration depends only on lift-to-drag ratio. Deceleration time is
\be
t-t_i=\fr{v_0(L/D)}{2g}\log\left(\fr{[1+v_i/v][1-v/v_0]}
{[1-v_i/v_0][1+v/v_0]}\right)\,.
\label{eq:a15}
\end{equation}

The heat generated per unit area per unit time can be found from
\be
\fr{Q}{A}=\fr{3\rho_0 v_i^3}{2}\left[\left(\fr{Re}{M d}\right)_0 \right]^{-1/2}
\sqrt{\g}\left(\fr{v}{v_i}\right)^3=\s T^4\,,
\label{eq:a16}
\end{equation}
where, $Re/M d$ is the Reynold's number per unit length $d$ per Mach number $M$ at
the surface, which is $\cong 5\times 10^5$ for terrestrial-size planets, and $\s$ is
the Stefan-Botzmann constant. The maximum
temperature reached on deceleration at a location, where $v/v_i = \me^{-1/6}=0.83$, i.e. 
$(\rho_0\g/2\mu)(C_{\rm D}A/m\sin{\phi})=1/6$, is
\be
T^4_{\rm max}=\fr{1}{\s}\left( \fr{Q}{A} \right)_{\rm max}=
\Big[ 0.4\rho_0 v_i^3(\sin{\phi})^{1/2}\Big] \left[ \left( \fr{Re}{M d} \right)_0
\fr{C_{\rm D}A}{m} \fr{\rho_0}{2\mu}\right]^{1/2}\,.
\label{eq:a17}
\end{equation}
For the example of an impactor with $m=10^{24}$ kg, $A=10^8$ km$^2$, $v_i=10$, 
$\phi=60^{\circ}$ and so on, we estimate $T_{\rm R} \approx 2 \times 10^4$ deg, 
and $L\approx 7 \times 10^{31}$ erg/sec for the hot spot with $R_{\rm E}$ radius.    
The deceleration time from Eq.~\ref{eq:a15} for this example will be of order 
$10^3$ sec.

The integrated energy released during descent will be
\be
\dder{(\rm Power)}{t}=3\rho A v^2 \dder{v}{t}=3\rho A v^2 g_{\rm J} \left(\fr{L}{D}+
\left[\mu r \fr{C_{\rm D}A}{m}\fr{\rho_0\g}{2\mu}\right]^{-1}\right)^{-1}\,
\label{eq:a18}
\end{equation}
which gives us  
\be
E=\int P dt\times A = \left[\left(\fr{9\pi}{2}\right)^{1/2}v_i^3
\left(\fr{\rho_0}{2\mu}\right)^{1/2}\right]\left[\left(\fr{Re}{M d}\right)_0
\fr{C_{\rm D}A}{m} \sin{\phi}\right]^{r/2}\,.
\label{eq:a19}
\end{equation}

\nocite{*}
\bibliographystyle{spr-mp-nameyear-cnd}
\bibliography{biblio-u1}

\begin{thebibliography}{}
\bibitem[Almoznino et al. 2005]{elhanan2005} Almoznino, E., Brosch, N., 
Shara, M., \& Zurek, D.\  2005 \mnras, 357, 645
\bibitem[Asphaug et al. 2006]{nature-paper} Asphaug, E., Agnor, Craig B., \& Williams, Q., 2006,
\nat, 439,155
\bibitem[Becker et al. 2004]{Becker} Becker, A.~C., et al. 2004, \apj, 611, 418
\bibitem[Brassart \& Luminet 2008] {luminet} Brassart, M. and Luminet, J.-P., 
2008, Astron. Astrophys. 481, 259-277
\bibitem[Brosch 1994]{noah-IAU168-94} Brosch, N.\ 1996,  
Proceedings from IAU Symposium 168, eds. Minas C. Kafatos, and Yoji Kondo, 
(Kluwer Academic Publishers, Dordrecht), 553
\bibitem[Brosch 1998a]{noah-phys-scripta98} Brosch, N. 1998a, {\it Physica Scripta}, T77, 16
\bibitem[1998b]{brosch-esa} Brosch, N. 1998b, ESASP, 413, 789
\bibitem[Chevalier \& Sarazin 1994]{sarazin} Chevalier, R.~A. \& Sarazin, C.~L., 1994,
\apj, 429, 863
\bibitem[Chyba et al. 1993]{chyba} Chyba, C.~F., Thomas, P.J., \& Zahnle, K.~J.
1993, \nat, 361, 40
\bibitem[Clowe et al. 2006]{bullet} Clowe, D., Brada\u{c}, 
M., Gonzalez, A.~H., Markevitch, M., Randall, S.~W., Jones, C., \& 
Zaritsky, D.\ 2006, \apjl, 648, L109 
\bibitem[Courtes et al. 1995]{FAUST} Courtes, G., Viton, M., 
Bowyer, S., Lampton, M., Sasseen, T.~P., \& Wu, X.-Y.\ 1995 \aap, 297, 338
\bibitem[Gezari et al. 2008]{GALEX-BH} Gezari, S., et al.\ 2008, \apj, 676, 944 
\bibitem[Gizis et al. 2002]{Gizis} Gizis, J., Reid, I.~N., and 
Hawley, S., 2002, \aj, 123, 3356
\bibitem[1975]{monochromatic} Hayes, D.~S., \& Latham, D.~W.\ 1975, \apj, 197, 593 
\bibitem[H{\'e}brard et al. 2004]{chtonian} H{\'e}brard, G.,
Lecavelier Des {\'E}tangs, A., Vidal-Madjar, A., D{\'e}sert, J.-M., \&
Ferlet, R.\ 2004, Extrasolar Planets: Today and Tomorrow, ASP Conf. Proc., 321, 203
\bibitem[Jee et al. 2007]{cl0024} Jee, M.~J., et al.\ 2007, \apj, 661, 728 
\bibitem[Kim et al. 1999]{kim} Kim, S.~S., Park, M.-G., 
\& Lee, H.~M.\ 1999, \apj, 519, 647 
\bibitem[Kulkarni \& Rau 2006]{DLS} Kulkarni, S.~R. and Rau, A. 2006, \apj, 644, L63
\bibitem[Martin et. al. 2005]{Martin2005} Martin, D.~C., et al.\ 2005, \apj, 619, L7
\bibitem[Mitra-Kraev et al. (2005)]{Mitra-Kraev} Mitra-Kraev, U., et 
al.\ 2005, \aap, 431, 679 
\bibitem[Parker et al. 2001]{UIT-parker} Parker, J.~W., Zaritsky, D., 
Stecher, T.~P., Harris, J., \& Massey, P.\ 2001 \aj, 121, 891
\bibitem[BASI June 2007]{BASI-tauvex} Proceedings of the TAUVEX March 2006 Science 
Meeting, eds. K.~Rao and M.~Safonova, 2007, Bull. Astron. Soc. India, 35, 169-300
\bibitem[Robinson et al. 2005]{galex-uvceti-flash} Robinson, R.~D. et al. 2005, \apj,  633, 447 
\bibitem[Safonova \& Almoznino 2007]{TOM} Safonova,~M. and Almoznino,~A. 2007, 
{\it The TAUVEX Observer's Manual}, v1.0 (http://tauvex.iiap.res.in)
\bibitem[Sivaram 2007]{sivaram} Sivaram, C.\ 2007, ArXiv e-prints, 707, arXiv:0707.1091 
\bibitem[1987]{smith} Smith, A.~M., Cornett, R.~H., \& Hill, R.~S.\ 1987, \apj, 320, 609 
\bibitem[Stern 1994]{stern} Stern, A. 1994, \apj, 108, 2312 
\bibitem[Stevenson 1987]{stevenson87} Stevenson,D.~J. 1987 {\it Ann. Rev. Earth Planet. Sci.}, 15, 271
\bibitem[Tamm \& Spruit 2001]{sivaram-ref4} Tamm,~R. \& Spruit, H., 2001, 
\apj, 561, 329; Chen, W. \& Li, X., 2006, \aap, 450, L1 
\bibitem[Turnbull \& Tarter 2003]{Turnbull} Turnbull, M.C. and Tarter, J., 2003,
\apj S, 145, 181 
\bibitem[van den Oord et al. 1996]{Oord} van den Oord, G.~H.~J., et al. 1996,
\aap, 310, 908 
\bibitem[Welsh et al. 2005]{galex-catalog} Welsh, B.~Y., et al.\ 2005, \aj, 130, 825 
\bibitem[Welsh et al. 2006]{galex-spectra} Welsh, B.~Y., et al.\ 2006, \aap, 458, 921 
\bibitem[Welsh et al. 2007]{galex-flares} Welsh, B.~Y., et al.\ 2007, \apjs, 173, 673 
\bibitem[Zhang \& Sigurdsson 2003]{zhang} Zhang, B. \& Sigurdsson, S.\  2003, \apjl, 596, L95 
\bibitem[Zwart et al. 1999]{sivaram-ref3} Zwart,~S., et al.\ 1999, \aap, 348, 117 

\end{thebibliography}

\end{document}